\documentclass[]{aastex63} 
\usepackage{csvsimple}
\usepackage{multirow,bigdelim}

\usepackage{amsmath}
\usepackage{amssymb}

\newcommand{\om}{`Oumuamua}

\usepackage{colortbl}
\definecolor{tablegray}{rgb}{0.89, 0.89, 0.89}

\begin{document}

\title{Shape-Driven Selection Effects for Aspherical Near-Earth Objects in Systematic Surveys}
\correspondingauthor{W. Garrett Levine}

\email{garrett.levine@yale.edu}

\author[0000-0002-1422-4430]{W. Garrett Levine}
\affil{Dept. of Astronomy, Yale University, 52 Hillhouse, New Haven, CT 06511, USA}

\author[0000-0001-7830-028X]{Robert Jedicke}
\affil{Institute for Astronomy, University of Hawaii, 2680 Woodlawn Dr, Honolulu, HI 96822, USA}

\begin{abstract}

The apparent magnitude of elongated small bodies is time-dependent over their rotation phase. Therefore, previously undiscovered aspherical minor planets may experience a shape-driven selection effect in systematic surveys versus their spherical counterparts. In this study, we conduct injection-recovery exercises of synthetic asteroid lightcurves using a simple model to quantify the effect of varying axial ratio on detection efficiencies. We find that high-amplitude lightcurves are confronted with adverse selection effects for survey cadences and discovery thresholds for constructing tracklets that are representative of modern and proposed future NEO searches. Furthermore, we illustrate the possible hazards of drawing population-level inferences on an underlying reservoir of elongated small bodies. If physical size and characteristic axial ratios are correlated, then size-frequency distributions may require revision at small diameters. In particular, this effect could alter the estimated populations of near-Earth objects. We conclude by discussing the applicability of our results to various other classes of solar system minor planets and interstellar interlopers, as well as discuss future work that may further interrogate this detection bias.

\end{abstract}


\section{Introduction} \label{sec:intro}

Debiasing observations in small body surveys to reveal the underlying population-level statistics is critical to understanding our solar system's inventory of comets and asteroids. A substantial body of literature has investigated the ability for observatories to identify asteroids while varying parameters such as size \citep{stuart2004biasPopulation,heinze2021neoPopulation}, color \citep{luu1989color}, viewing geometry \citep{lu2019elongated}, orbit \citep{bottke2000NEOscience,bottke2002neo, jedicke2016selection, granvik2018NEOdebiased, nesvornyInPrep}, and survey photometric efficiency \citep{veres2017lsstSimulation}, among other factors. Deducing the population of near-Earth objects (NEOs) brings scientific benefits, is important for planetary defense, and could have commercial applications.

Due to their proximity to Earth, NEOs --- often defined as minor planets with perihelia less than $1.3\,\text{au}$ \citep{bottke2000NEOscience} --- can be discovered at smaller physical sizes than Main Belt Asteroids (MBAs), Trans-Neptunian Objects (TNOs) and other populations of minor planets. Asteroids with surface gravities that do not overwhelm their material strength (typically with diameters $d < 200\,\text{km}$) can be aspherical, leading to variable lightcurves and correspondent selection effects in surveys. Elongated asteroids with apparent magnitudes near a survey's limit may change in brightness and vary between being detectable and non-detectable while spinning. Stated differently, the search volume to which a survey is sensitive to a non-spherical small body is time-dependent. Because NEO surveys differ in sky coverage, limiting magnitude, cadence, and detection thresholds, the repercussions of shape-dependent variability depend on the observatory and discovery strategy. As surveys push towards fainter limiting magnitudes, however, the trend towards discovering smaller asteroids will continue. Thus, accounting for shape-driven variability will be increasingly important if lightcurve amplitude is correlated with size.

Previously, shape-driven selection effects were mentioned by \cite{lu2019elongated} in a study that computed lightcurves of ellipsoidal asteroids at varying phase angles. As detection efficiencies were not the main point of that paper, however, the consequences of small body axial ratios on survey results were not thoroughly examined. Other simulation-based studies have also acknowledged that elongated minor planets may experience differing detection probabilities versus their spherical counterparts \citep{veres2017lsstSimulation}, but have not provided a detailed account of this bias. \citet{masiero2009thousand}'s observational survey provided debiased distributions of axial ratios and rotation periods for MBAs, but implicit in their procedures was the assumption that all shapes were identified with the same efficiency by the the Moving Object Processing System (MOPS) discovery pipeline \citep{denneau2013MOPS}.

The scientific literature currently lacks a complete and focused evaluation of high-amplitude lightcurve selection effects. Here, we strive to provide this necessary research. \citet{navarroMeza2021lightcurveBias} examined shape-driven variability, but only in the case of objects that were nominally dimmer than the system limiting magnitude. Furthermore, their assumed observing strategy was more applicable to MBAs. In this study, we consider NEOs that are both brighter and dimmer than the system's limiting magnitude, examine asymmetric lightcurves, and illustrate the population-level implications of shape-driven variability.

Beyond the scope of NEOs, biases in detecting ellipsoidal small bodies have ramifications for our understanding of the Milky Way's reservoir of interstellar objects (ISOs). For example, 1I/\om's lightcurve varied quasi-periodically by approximately three magnitudes over a duty cycle of several hours and immediately indicated a more extreme shape than any solar system analog \citep{Meech2017}. Later, \cite{mashchenko2019modelling} demonstrated that a 6:6:1 ellipsoid with some tumbling and torques was the best-fit.

Although a number of hypotheses have been proposed to explain \om's shape and other curious properties, including an eroded hydrogen iceberg \citep{fuglistaler2018solid, Seligman2020, levine2021assessing}, nitrogen iceberg \citep{jackson20211i, desch20211i}, ultra-porous dust aggregate \citep{moro2019could, luu2020oumuamua}, and an artificial object \citep{bialy2018could}, none of these interpretations simultaneously satisfy the theoretical constraints and observational data \citep{jewitt2022ISOreview}. For example, one aspect of \om{} that must be explained by any viable formation hypothesis is the Galactic population of similar ISOs. Detailed analyses of the Pan-STARRS survey volume \citep{Engelhardt2017, Meech2017, Do2018} yielded spatial number densities of $n_{\text{iso}} \sim 0.1\,\text{au}^{-3}$ for \om-like objects in the inner solar system. Those studies did not account for \om's elongation, so those reported values may require revision if \om{} was more or less detectable due to its shape. Correspondingly, a modified ISO reservoir may affect the interpretations of \om's composition that are theoretically viable.

In this study, we quantify the effect of shape-driven lightcurve variability on the detection probabilities for elongated small bodies in systematic surveys. Considering asteroids as triaxial ellipsoids and implementing observing strategies that are typical of NEO searches, Section \ref{sec:model} introduces a simulated survey model by which to isolate the effect of lightcurve variability. Then, Section \ref{sec:results} presents results on the sensitivity of these mock surveys to asteroids of particular shapes, rotation periods, and average apparent magnitude. Next, Section \ref{sec:population} illustrates potential failure modes of standard debiasing frameworks: accounting for shape-driven selection effects in shape distributions and size-frequency distributions (SFDs). Afterwards, Section \ref{sec:discussion} provides commentary on our results and elucidates the ramifications of our findings on future efforts to identify small bodies. Finally, we summarize our findings in Section \ref{sec:conclusion}.

\section{Modeling Shape-Driven Variability} \label{sec:model}

We analytically idealize asteroid lightcurves and consider fiducial survey parameters that are representative of modern NEO searches. Moreover, we restrict the allowable degrees of freedom for our synthetic small bodies in order to discern shape-driven selection effects in our simulations.

\begin{figure}
\epsscale{1.2}
\plotone{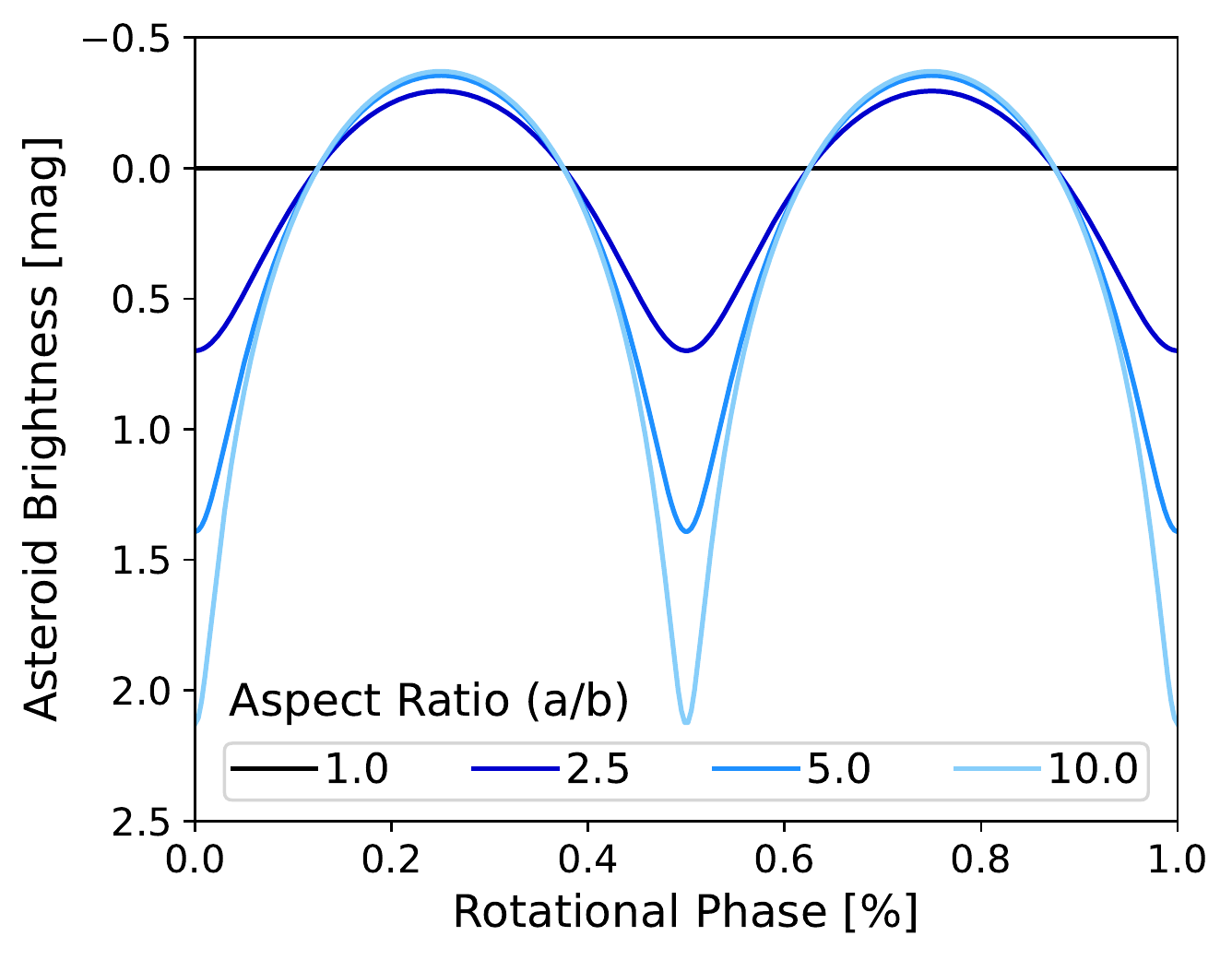}
\caption{Model lightcurves of asteroids computed from Equation \ref{eq:lightcurve} such that the median magnitude is zero over one full rotation. The aspect ratio $(a/b)$ is given in the legend; variability becomes more extreme for elongated objects.}
\label{fig:elongated}
\end{figure}

\subsection{Constructing Synthetic Lightcurves}

The asteroids are modelled as triaxial ellipsoids with semiaxes $a > b > c$ that rotate about their $c$ axes and have uniform surface albedos. We assume that observations occur at zero solar phase angle and that the asteroids' spin vectors are orthogonal to the line-of-sight such that the $c$ axis does not affect the lightcurve. NEO viewing geometry is often more extreme than our prescription, but considering consistent setups for our simulations reduces the number of variables for which our numerical experiments must control. Additionally, we assume geometric scattering; the asteroid's brightness scales linearly with the visible surface area as viewed from Earth and illuminated by the Sun. With this fiducial setup, the time-dependent projected area is \citep{detal1994lightcurve}

\begin{equation}
    S(\phi) \propto \sqrt{\frac{\sin^{2}{(\phi)}}{a^{2}} + \frac{\cos^{2}{(\phi)}}{b^{2}}}\,,
\end{equation}

\noindent where the instantaneous rotation phase is $\phi = (2\pi t/P) + \phi_{0}$, with the timestamp at the observation epoch $t$, the asteroid's rotation period $P$, and the initial rotation phase $\phi_{0}$. Therefore, the time-dependent lightcurve brightness in magnitudes can be expressed as

\begin{equation} \label{eq:lightcurve}
    m(t) = -2.5\log_{10}\Big(S(\phi)\Big)\,.
\end{equation}

While more detailed analytic \citep{barucci1982analyticLightcurve, connelly1984ellipsoidLightcurve, muinonen2015lightcurve}, statistical \citep{muinonen1998gaussianShape}, numerical \citep{lu2019elongated, taylor2022tidal} and ray-tracing models \citep{kaasalainen2001inversion, Seligman2019} exist, including ones that consider non-ellipsoidal configurations, realistic mineralogy, and varied observing angles, we opt for Equation \ref{eq:lightcurve} to cleanly isolate the shape-driven selection effects. Placing the synthetic asteroids at opposition renders many of these morphological, compositional, and geometric considerations inconsequential for our results. 

Figure \ref{fig:elongated} displays example lightcurves from applying Equation \ref{eq:lightcurve} with various axial ratios. The functions are double-peaked due to their symmetric shapes, and the peak-to-trough brightness change is

\begin{equation} \label{eq:amplitude}
    |\Delta m| =|2.5 \log_{10}(a/b)|\,.
\end{equation}

For a given axial ratio, our choices of phase angle and spin axis orientation minimize and maximize the lightcurve amplitude, respectively. As such, we will display many of our results in terms of both $(a/b)$ and the corresponding $|\Delta m|$. We also ignore modifications to lightcurves from non-principal axis rotation \citep{Rafikov2018}, binary asteroids \citep{seligman2021spinChaos}, or sublimation-induced spin-up \citep{mashchenko2019modelling, taylor2022tidal}. These situations only apply to some NEOs and are unimportant at the population-level, but determining the detection probability of a specific small body could require a consideration of these effects.

\subsection{Simulating Small Body Surveys}

NEOs are most likely to be detected near the survey system's limiting magnitude due to their steep SFD. If we assume that the efficiency for identifying an asteroid is 100\% for objects with instantaneous apparent magnitude $m$ less than the system's limiting magnitude $m_{lim}$ and define $m_r \equiv m - m_{lim}$, then objects are detected if $m_r < 0$. Here, $m_r$ is the small body's instantaneous apparent magnitude relative to the survey's limit. Working in terms of $m_r$ means that our formalism can conveniently model asteroids of different sizes within the same numerical framework.

A better representation of the detection efficiency accounts for image noise in its various forms, which we model as

\begin{equation}\label{eq:probability}
    p(t) = p_{max}\bigg(1 + \exp{\bigg(\frac{m_r(t)}{w}\bigg)}\bigg)^{-1}\,,
\end{equation}

\noindent where $p_{max}$ is sometimes called the ``fill factor" \citep{veres2017lsstSimulation} and accounts for chip gaps and other detector imperfections. In Equation \ref{eq:probability}, $w$ parameterizes the magnitude range over which the detection efficiency declines quickly at the system limiting magnitude. Our Equation \ref{eq:probability} for per image detection probability is of the same form as Equation 15 in \cite{jedicke2016selection}, who applied the equation to single-night tracklet detection efficiency. In small body surveys, a tracklet is composed of detections from individual images that are associated with the same object. We assume $p_{max} \sim 1$ and $w = 0.25$ \citep{Chambers2016-PS1}, which correspond to a minimum signal-to-noise for a single detection of $\text{SNR} \sim 4 \sim (1/w)$.

To examine the effect of cadence and discovery criteria on shape-driven selection effects, we simulated six different nightly survey scenarios (Table \ref{tab:cadenceTable}) encompassing a representative set of parameters for modern NEO searches. Specifically, Survey~2 is reminiscent of nightly strategies from prolific surveys from the 2000s and 2010s: the Catalina Sky Survey and Pan-STARRS \citep[][\ respectively]{Larson1998,Chambers2016-PS1}.  These surveys linked the detections from one night into `tracklets' that were then submitted for followup by other observers.  If the tracklets were confirmed, then the survey could claim a ``discovery." In contrast, Survey~5 corresponds to a nominal cadence for the upcoming Large Synoptic Survey of Space and Time (``the LSST"), with only two visits to a particular area of the sky on one night \citep{ivezic2019lsst}. These two-detection tracklets will be too faint for followup by most other observers and probably have a high false-positive rate, so the LSST will link detections across multiple nights before reporting an object to the Minor Planet Center \citep{jones2009lsstSolarSystem} We include Surveys~1, 3, 4, and 6 to extend the parameter space of cadence and tracklet-building criteria which we examine.

\begin{figure}
\epsscale{1.15}
\plotone{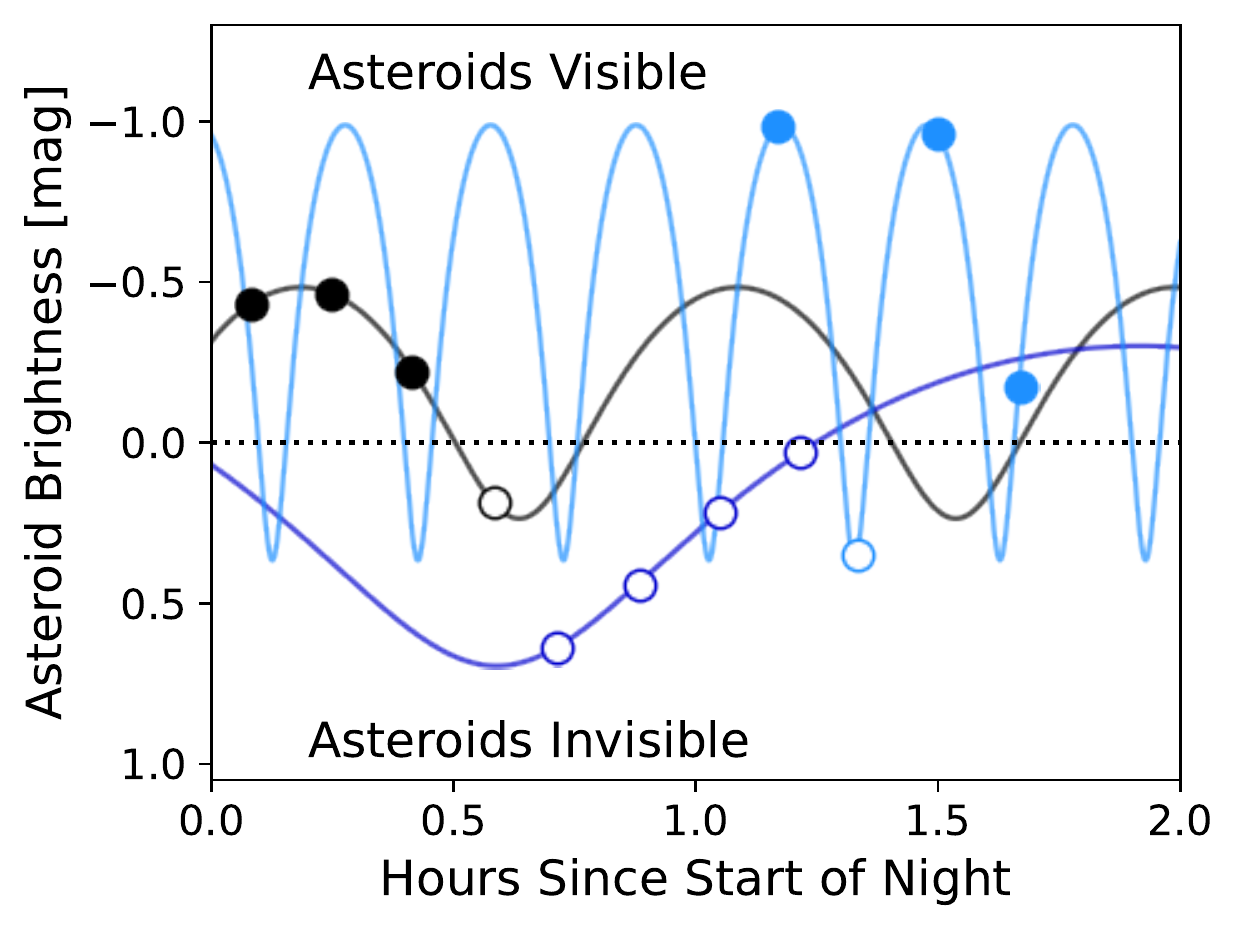}
\caption{Synthetic asteroid lightcurves, shown in magnitudes as $m_r \equiv m - m_{lim}$, versus time. Illustrated are simulated exposures corresponding to the cadence of Survey~1 \& 2, where the circular markers correspond to $m_r$ at the epoch of observation. Filled dots denote frames where the asteroids were visible, and unfilled dots indicate that the asteroid was invisible at the epoch of observation. The simulated lightcurves were initialized according to Equation \ref{eq:lightcurve} with the following parameters for the black, dark blue, and light blue curves, respectively: periods $P = \{1.8\,\text{h}, 5.3\,\text{h}, 0.6\,\text{h}\}$, axial ratios $(a/b) = \{1.94, 2.50, 3.48\}$, and median apparent magnitudes $m_r = \{-0.23, 0.00, -0.65\}$.}
\label{fig:modelDetections}
\end{figure}

To simulate observing a given small body, we compute $m_r$ at each exposure time based on $m$ given by Equation \ref{eq:lightcurve} and the object's assigned $P$, $(a/b)$, and $\phi_{0}$. Next, we compute the probability of detection $p$ for each individual frame from Equation \ref{eq:probability} and draw a random number $q$ on the uniform $[0, 1]$ interval. We consider the asteroid to be visible in a given frame if $p > q$ and take all observations to be conducted in the same photometric band.  Finally, we require that the collection of exposures passes the detection criterion in Table \ref{tab:cadenceTable} for that asteroid to be counted as ``identified." We ignore any variable background brightness due to the Moon's phase, missing observations due to inclement weather, or other factors that are not directly related to the asteroid's shape-driven lightcurve variability. Figure \ref{fig:modelDetections} illustrates our simulations applied to attempted detections of three elongated objects by the ten-minute cadence of Survey~1 \& 2. The former survey would not have successfully identified any tracklets, while the latter survey would have identified two asteroids.

We work in the limit where the exposure time is shorter than meaningful changes in $m_r$. Since exposure times are typically less than one minute and $P$ for diameter $d \gtrsim 200\,\text{m}$ objects is typically on the order of hours \citep{warner2009database}, $m_r$ for these asteroids should change negligibly while the telescope aperture is open. Rotation periods can approach the exposure times of contemporary surveys and the future LSST cadence for decameter-scale objects \citep{Farinella1998-MeteoriteDeliveryViaYarko, Bolin2014-Minimoons}, so obtaining robust $P$ constraints for those asteroids requires targeted followup \citep{beniyama2022video}. Without such efforts, extremely fast-rotators ($P \ll 60\,\text{s}$) are likely to have unidentified periods. Correcting for these selection effects will require additional effort. For the purposes of characterizing shape-driven variability, pipeline-specific effects that are incorporated into efficiency functions via flat-fielding or linking across fields are unimportant and can be ignored.

We assume that a given asteroid is identifiable for only one night and that each survey in Table \ref{tab:cadenceTable} operates independently. Small bodies may actually be found by any number of programs that can have overlapping discovery parameter space, but we are concerned with the ability of a given asteroid detection pipeline to construct tracklets that can be targeted by other followup systems for confirmation, photometry, and spectroscopy. Objects that venture closer to Earth and become brighter than $m_{lim}$ for all $\phi$ do not experience shape-driven selection effects. Thus, we are primarily interested in asteroids whose trajectories graze the survey's effective volume and allow just a one-time opportunity for discovery. \cite{nesvornyInPrep} found that small NEOs in the Catalina Sky Survey were almost always detected on only one night, thereby justifying our model.

\begin{table}[]
\centering
\begin{tabular}{|ccc|}
\hline
\multicolumn{3}{|c|}{\textbf{Simulated Nightly Surveys}} \\ \hline
\multicolumn{1}{|c|}{\textbf{Label}} & \multicolumn{1}{c|}{\textbf{Cadence {[}min{]}}} & \multicolumn{1}{c|}{\textbf{n/m Tracklet}} \\ \hline
\multicolumn{1}{|c|}{Survey 1} & \multicolumn{1}{c|}{10 Minutes} & \multicolumn{1}{c|}{4/4} \\ \hline
\multicolumn{1}{|c|}{Survey 2} & \multicolumn{1}{c|}{10 Minutes} & \multicolumn{1}{c|}{3/4} \\ \hline
\multicolumn{1}{|c|}{Survey 3} & \multicolumn{1}{c|}{30 Minutes} & \multicolumn{1}{c|}{4/4} \\ \hline
\multicolumn{1}{|c|}{Survey 4} & \multicolumn{1}{c|}{30 Minutes} & \multicolumn{1}{c|}{3/4} \\ \hline
\multicolumn{1}{|c|}{Survey 5} & \multicolumn{1}{c|}{15 Minutes} & \multicolumn{1}{c|}{2/2} \\ \hline
\multicolumn{1}{|c|}{Survey 6} & \multicolumn{1}{c|}{15 Minutes} & \multicolumn{1}{c|}{2/3} \\ \hline
\end{tabular}
\caption{Characteristics of simulated surveys that were applied to synthetic asteroid populations. ``Cadence" refers to the time between exposures. For example, the ten-minute cadence for Survey~1 and Survey~2 means that frames were obtained at $t = \{0, 10, 20, 30\}\,\text{min}$. ``n/m Tracklet" corresponds to the threshold for creating a tracklet on a given} night. For example, the 3/4 on Survey~2 and Survey~4 states that asteroids must be detected in three out of four frames to form a tracklet.
\label{tab:cadenceTable}
\end{table}

\section{Single-Object Selection Effects} \label{sec:results}

Here, we apply our survey and lightcurve models from Section \ref{sec:model} to synthetic NEOs. We assess detection efficiencies as a function of the asteroids' median $m_r$ over one rotation period. Normalizing to the mean flux would not be a sensible choice since detection is a binary event; an object is either visible or invisible in a given frame.

\subsection{Asteroid Detection Efficiencies with Axial Ratio}

First, we examine the impact of aspect ratio on detection probabilities while varying average relative brightness compared to the survey limit. We run each fiducial survey from Table \ref{tab:cadenceTable} on a set of $10^{7}$ synthetic small bodies with median $m_r \in [-1.5, 0.5]$ and axial ratios $(a/b) \in [1.0, 4.0]$, where each range is divided into 100 equally-sized bins. In each grid cell, we initialize 1000 asteroids with $m_r$ and $(a/b)$ selected randomly from the possible values within that cell. Initially, all asteroids are assigned random $P \in [2, 8]\,\text{hr}$ and random $\phi \in [0, 2\pi]$. These $P$ are representative for asteroids larger than $200\,\text{m}$ \citep{pravec2000asteroidrotation}, but we will show later that our conclusions remain qualitatively unchanged for different rotation rate regimes.

\begin{figure*}
\epsscale{1.2}
\plotone{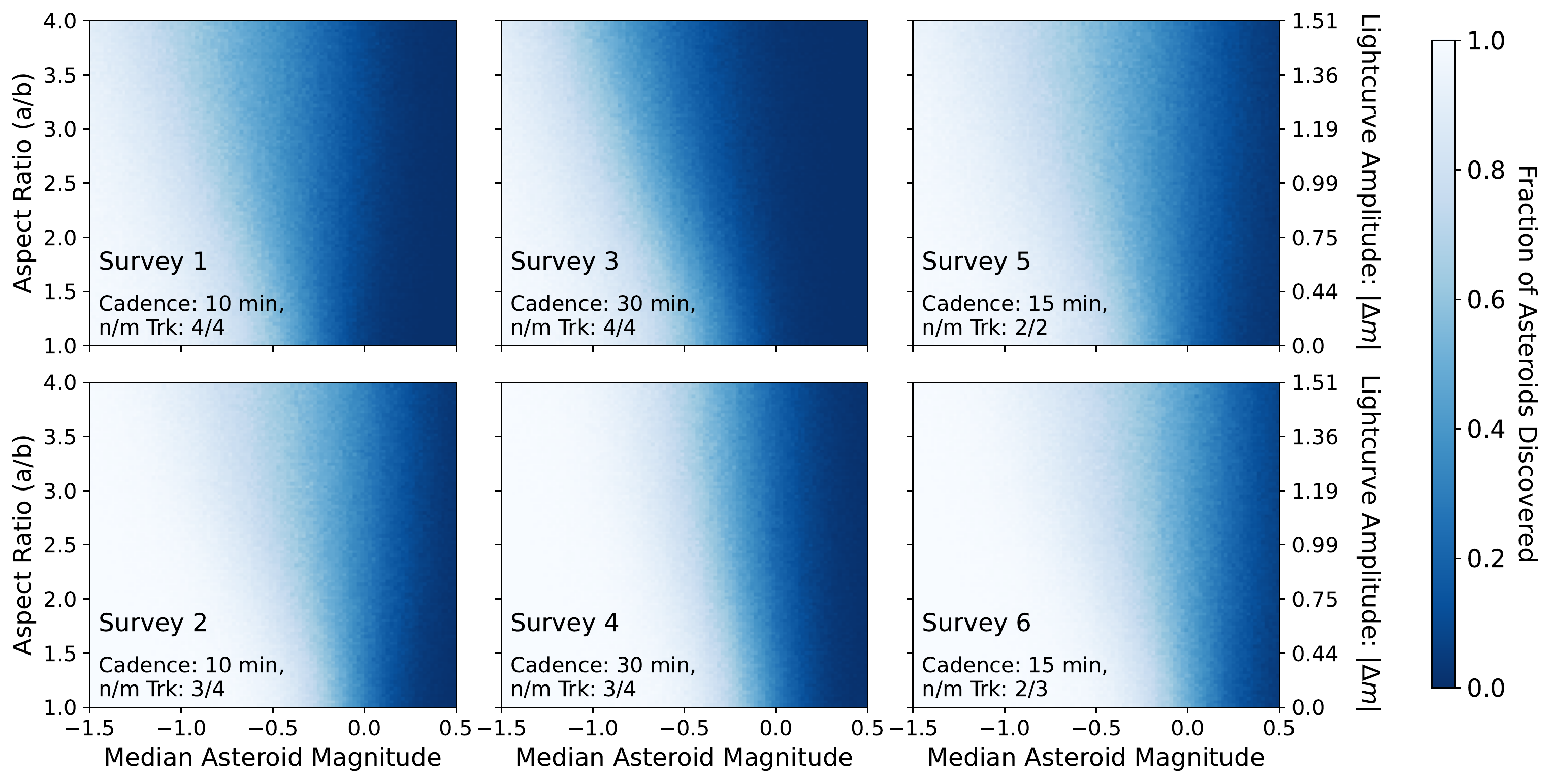}
\caption{Detection statistics from our injection-recovery model, displayed as 100 x 100 grids of aspect ratio ($a/b$) versus median asteroid magnitude for each simulated survey. Every cell represents the search for 1000 synthetic NEOs. Each heatmap displays the survey number and parameters as given in Table \ref{tab:cadenceTable}. The lightcurve amplitude, $|\Delta m|$, is given on the secondary (right-hand side) y-axis.}
\label{fig:aspectVsMag}
\end{figure*}

After generating the lightcurves according to Equation \ref{eq:lightcurve}, we imposed the observation cadences for each survey and evaluated whether the asteroids were identified according to the discovery criteria in Table \ref{tab:cadenceTable}. From these results, we computed the detection rate for each $(a/b, m_r)$ combination in the heatmaps on Figure \ref{fig:aspectVsMag}. A bias exists against detecting elongated objects and is strongest for Survey 3. For all surveys in Table \ref{tab:cadenceTable}, asteroids with $(a/b) \sim 3$ like that of the NEO 433 Eros must be more than 0.5 magnitudes brighter than the nominal limit for tracklets to be reliably created. The diagonal shape of the dark region, indicating the parameter space where asteroids are increasingly undiscovered, shows that extreme ellipsoids can be routinely missed even if $m_r < 0$ because the minima of Equation \ref{eq:lightcurve} are increasingly dimmer with increasing axial ratio. Notably, the survey efficiency does not drop immediately when $m_r \sim |\Delta m|$. Instead, the detection probabilities drop only as a substantial fraction of the lightcurve becomes dimmer than $m_{lim}$. This behavior is related to the broad maxima and narrow minima that emerge from Equation \ref{eq:lightcurve}.

The boundary between detectable and undetectable parameter space is not sharply delineated in Figure \ref{fig:aspectVsMag}. Indeed, the $w$ parameter dictated by the photometric noise at $m_{lim}$ alleviates some shape-driven biases by introducing more randomness into discovery probabilities. Whether an asteroid is visible at a given epoch depends not only on the deterministic rotational phase and median $m_r$, but also on a probabilistic component from the low $\text{SNR}$ regime. Increasing $w$ will increase the importance of stochasticity at the expense of lightcurve amplitude. If a higher $\text{SNR}$ cutoff were chosen, then the correspondingly lower $w$ would make the trends in Figure \ref{fig:aspectVsMag} sharper but would decrease the overall survey discovery efficiency. A lower $p_{max}$ in Equation \ref{eq:probability} would also decrease the magnitude of shape-driven selection effects, but decreasing $p_{max}$ is undesirable because that would reduce the survey efficiency for brighter objects.

The effect of $w$ also explains why spherical objects of zero lightcurve amplitude are more difficult to detect with the stricter detection thresholds in the odd-numbered surveys. Evaluating Equation \ref{eq:probability} for $m_r = 0$ objects gives $p = 0.5$ in a given frame. In Survey 3, which requires 4/4 visible points for a successful discovery, the tracklet-building probability is $1/16 \sim 0.06$. For Survey 4 that requires 3/4 points, this discovery fraction is $5/16 \sim 0.31$. As $|\Delta m|$ exceeds $w$, however, shape-driven effects dominate the detection probabilities.

Despite the similar overarching trend for all surveys, we find noticeable differences in shape-driven selection effects with cadence and discovery criteria. Even mildly aspherical configurations with $m_r \sim 0$ may pose difficulties for the stringent detection criteria in Surveys~1 \& 3. Our even-numbered surveys, with more relaxed tracklet-building thresholds, successfully found more high-amplitude lightcurves than their odd-numbered counterparts. The discovery criterion most dramatically affects the longer-cadence simulations. When the single-night data collection epochs span a larger fraction of the rotation period, exposures are more likely to occur during both the brightness maxima and minima. As a consequence, the stringent discovery criterion in Survey~3 eliminates many objects that are found by the more lenient Survey~4. Encouragingly, the LSST-like cadence in Survey~5 identifies a larger fraction of elongated objects than either Survey~1 or Survey~3. Moreover, the small difference between Surveys~5 \& 6 demonstrates the minor marginal benefit of a third visit.

Taken together, our results show that selection effects against ellipsoidal small bodies objects can be problematic for NEO searches. More objects that are nominally brighter than $m_{lim}$ are missed than vice versa. The net change in overall survey yield due to shape-driven variability, however, depends on the interplay between the underlying SFD and shape distribution near the survey's limiting magnitude. In summary, asteroids are more likely to be undetected by sparse, non-targeted photometry as axial ratio increases. Coincidentally, these extreme small bodies often engender the most intrigue from the scientific community. Furthermore, our results suggest that planetary defense initiatives may be impaired by shape-driven NEO selection effects.

\subsection{Shape-Driven Variability and Absolute Magnitudes}

Because elongated objects near $m_r = 0$ are mostly detected near their brightness maxima, the derived diameter of elongated objects that are discovered by surveys will be skewed towards larger sizes in the absence of deep followup or careful debiasing. To interrogate the expected difference in $H$ from shape-driven variability, we ran Survey 3 on a fiducial set of small bodies. We used the same $P$ and $\phi$ distributions, the same number of grid cells, the same number of objects in each grid cell, and the same $(a/b)$ and $m_r$ range for the asteroids as we did to generate Figure \ref{fig:aspectVsMag}.

\begin{figure}
\epsscale{1.2}
\plotone{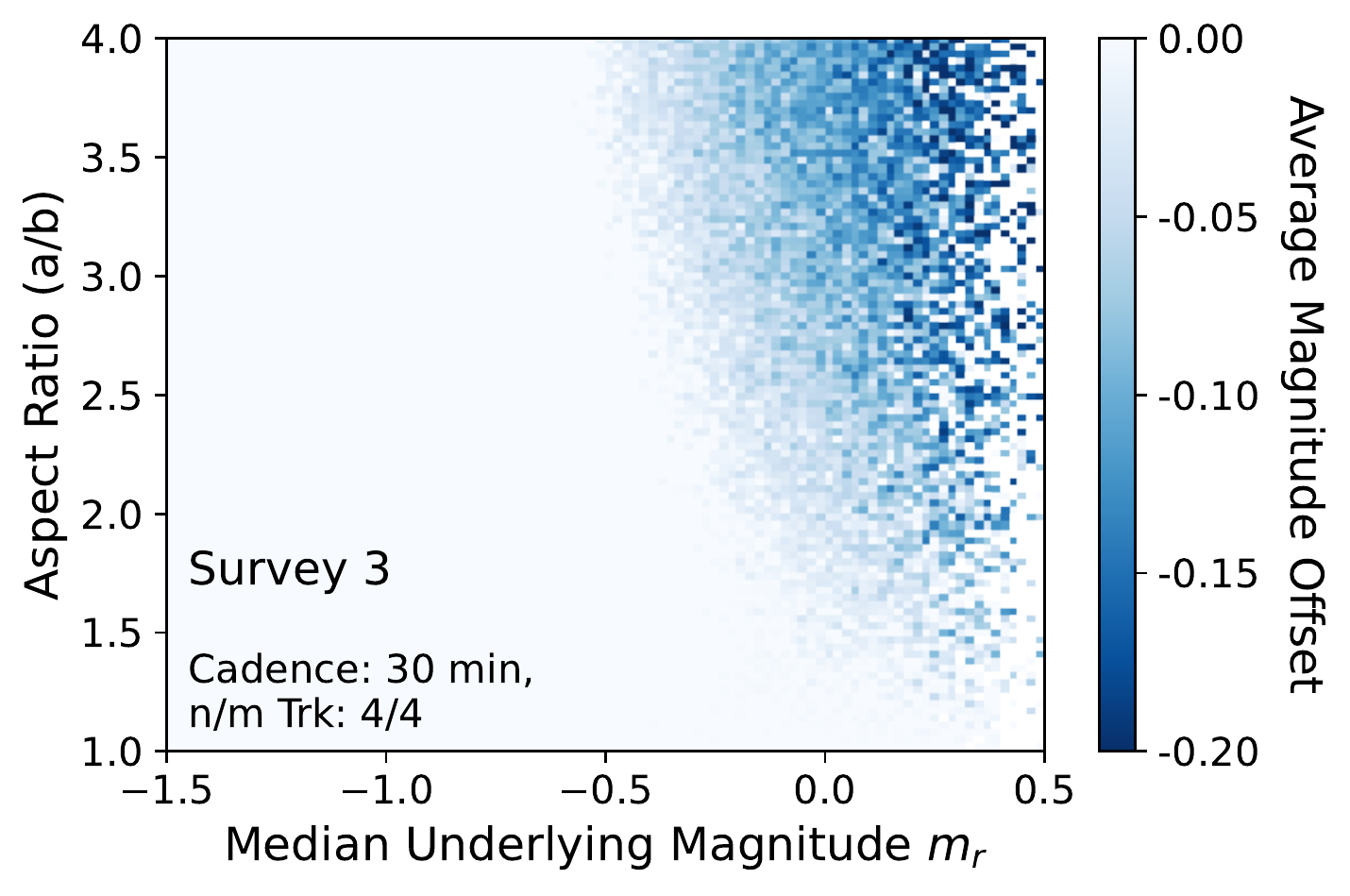}
\caption{Magnitude biases, calculated as $\Bar{m} - m_{r}$ for the asteroids that were detected by Survey~3 on a 100 x 100 grid of $(a/b)$ versus underlying $m_r$. Each grid cell corresponds to the search for 1000 synthetic NEOs. More negative values (darker colors) on the heatmap correspond to asteroids for which the recorded size is likely to be overestimate (i.e. the $H$ magnitude is underestimated). The blank, white cells on the right-hand side of the figure correspond to parameter space where no elongated objects were detected by Survey~3.}
\label{fig:Hmag}
\end{figure}

For each of the objects that were detected, we calculated the average apparent magnitude $\Bar{m}$ as the mean of the instantaneous asteroid magnitudes at each of the times that the object was successfully detected. Next, we subtract this value from the underlying median $m_r$. As with our previous methods, all brightness values are relative to $m_{lim}$. Our calculation gives the approximate error introduced by calculating $H$ from a survey's single-night positive detections without a high SNR followup characterization of the underlying lightcurve. Using our sign convention, grid cells with negative values correspond to asteroids with underestimated $H$ (i.e. a brighter calculated absolute magnitude than the actual value). The results (Figure \ref{fig:Hmag}) show that NEOs near $m_r = 0$ with large axial ratios tend to more severely underestimate $H$. Even in these extreme cases, the underestimates in $H$ (and overestimates in physical size) are smaller than $0.2\,\text{mag}$. In nearly all cases, the offset in $H$ is a factor of 2-3 lower than the underlying lightcurve amplitude. For asteroids with axial ratio $(a/b) < 2$, the errors in $H$ are lower. In regions of parameter space where few objects are identified, small number statistics from only taking the ``discovered" objects leads to scatter in Figure \ref{fig:Hmag}.

Recall that we calculated each asteroid's apparent magnitude at the epoch of observation without incorporating statistical fluctuations in the apparent flux. These fluctuations can be important near $m_{lim}$ since objects that are fainter than the limit can ``suffer" a random photometric bump into the regime with $m<m_{lim}$. Our method is not self-consistent because the apparent magnitudes of detected asteroids are always brighter than $m_{lim}$. In our simulations, asteroids with median $m_r < 0$ could occasionally have overestimated $H$; these objects could be serendipitously identified due to the survey's rapid but not-instantaneous drop in detection efficiency near $m_{lim}$ in spite of their shape-driven effects. We simply limit the grid cells at $\Bar{m} - m_{r} < 0$ to resolve this lack of fidelity. Future efforts to characterize shape-driven variability in specific surveys like LSST could more realistically incorporate the recorded $m_r$ near $m_{lim}$ into a debiasing procedure. For the purpose of this study --- to characterize the general behavior of selection effects related to axial ratio --- such a consideration of the photometric errors is unwarranted.

\subsection{Detection Efficiencies and Rotation Periods}

Here, we examine the relationship between $P$ and shape-driven selection effects for elongated small bodies over a broader range of periods. Considerable effort has already been put into debiasing $P$ distributions of asteroids \citep{masiero2009thousand, chang2015lightcurveSurvey}, as spin states can provide insight on internal structure for individual minor planets along with aggregate trends. Past studies have only considered biases related already-discovered asteroids. For example, rotation rates that are either much faster or much slower than the survey's sampling cadence are difficult to detect. None of this previous research considered the fact that shape-driven selection effects may inhibit the discovery of elongated objects with certain $P$.

\begin{figure}
\epsscale{1.2}
\plotone{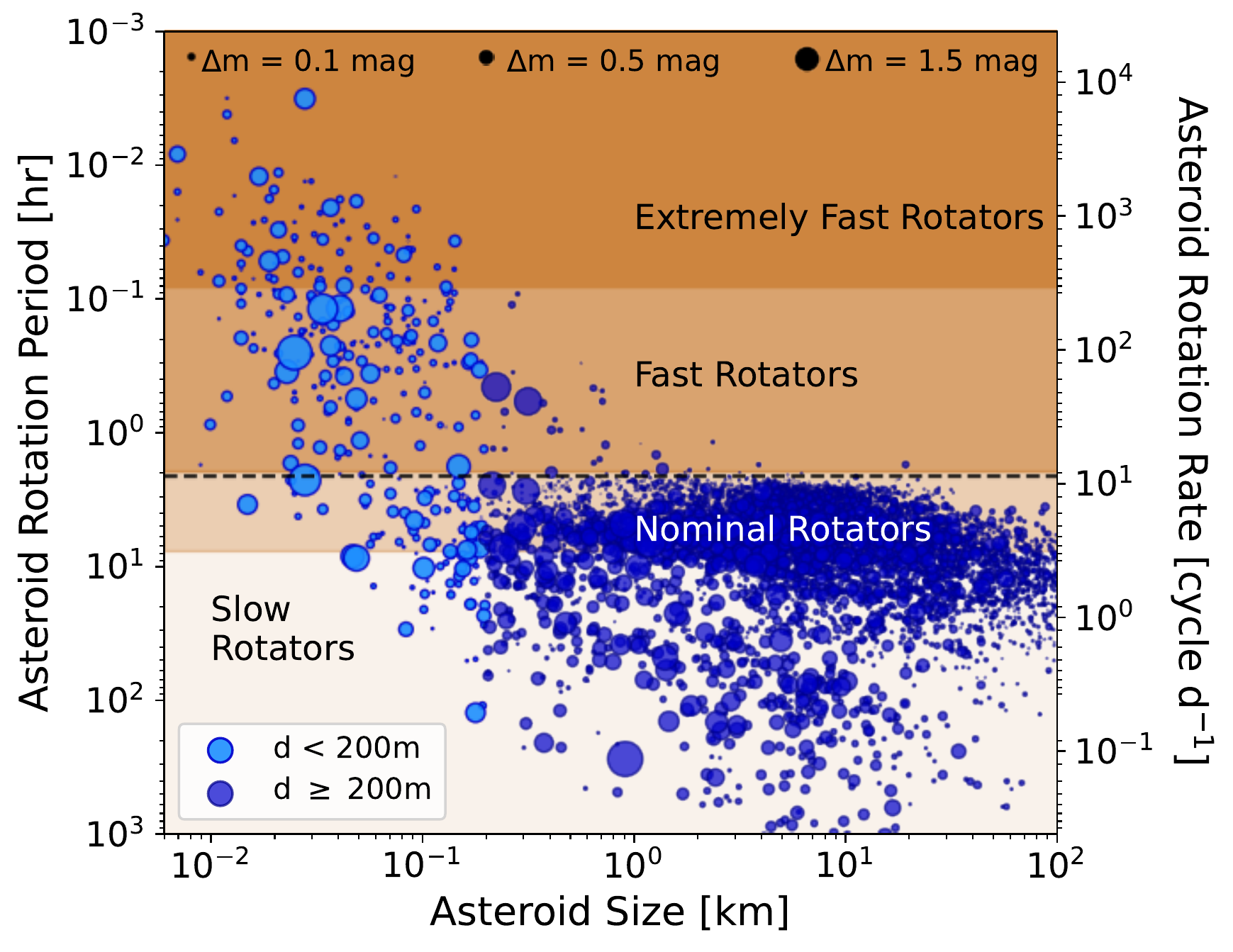}
\caption{Rotation period versus diameter for all objects in the Minor Planet Lightcurve Database with a reliability rating greater than 2; this set includes 2+, 2-, and 3 ratings. Marker size is reflective of the lightcurve amplitude, with larger markers indicating larger amplitude lightcurves. Asteroids are classified as either $d < 200\,\text{m}$ or $d \geq 200\,\text{m}$ via the marker color, and the background shading corresponds to the classification of $P$. The critical spin barrier for rubble piles $P_{c} \sim 2.2\,\text{h}$ is indicated by the black dashed line. Values of diameter reported in the lightcurve database are derived from numerous methods: indirectly from $H$ and estimated albedo and directly radar measurements, stellar occultations, or resolved images, among other methods.}
\label{fig:LCDB}
\end{figure}

We begin our rotation rate investigation by considering the classic scatter plot of $P$ versus physical size of objects from the Minor Planet Lightcurve Database (LCDB)\footnote{\url{https://minplanobs.org/mpinfo/php/lcdb.php}} \citep{warner2009database} on Figure \ref{fig:LCDB}. Asteroids with diameters $d \geq 200\,\text{m}$ usually rotate more slowly than the ``spin barrier" for the nominal stability of rubble piles \citep{pravec2000asteroidrotation, pravec2002asteroidsiii, pravec2006NEArotation}, indicating that material strength does not sculpt this population. The ``critical" rotation period at the spin barrier is given by

\begin{equation}
    P_{c} = 3.3\,\mathrm{h}\,\bigg(\frac{\rho}{1\,\mathrm{g}\,\mathrm{cm}^{-3}}\bigg)^{-1/2}\,,
\end{equation}

\noindent where $\rho$ is the bulk density of the rubble pile \citep{pravec2000asteroidrotation}. The spin barrier is often quoted as $P_{c} \sim 2.2\,\text{h}$ for S-type asteroids but can increase to $P\sim 3\,\text{h}$ for less dense C-types \citep{carbognani2017spinBarrier}.

To illustrate the interplay between $|\Delta m|$ and $P$ on detection fractions, we consider the following populations:

\begin{itemize}
    \item Extremely Fast-Rotators: $P < 5\,\text{min}$
    \item Fast-Rotators: $5\,\text{min} < P < 2\,\text{h}$
    \item Nominal Rotators: $2\,\text{h} < P < 8\,\text{hr}$
    \item Slow-Rotators: $P > 8\,\text{h}$
\end{itemize}

We define ``nominal rotators" with this terminology because $\sim58\%$ of the LCDB objects fall within this regime. Importantly, the LCDB set is biased against asteroids in the other groups. On the one hand, extremely fast-rotators require short cadences to sample above the Nyquist frequency. Without specialized efforts to target this spin regime, $P$ will be difficult to constrain. Indeed, \cite{beniyama2022video} recently found rotation rates as low as $P = 3\,\text{s}$ for $d \sim 20\,\text{m}$ NEOs by sampling lightcurves in 0.5s intervals. On the other hand, slow-rotators might not be observed for a sufficient baseline to resolve rotation periods. \cite{pal2020tess} analyzed known solar system minor planets that were serendipitously observed by the \textit{Transiting Exoplanet Survey Satellite} (\textit{TESS}), finding that the number of slow-rotators was previously underestimated due to this observational bias. Future studies with high cadence and long baseline are needed to constrain the distribution of the fastest and slowest rotating asteroids, respectively.

\begin{figure}
\epsscale{1.3}
\plotone{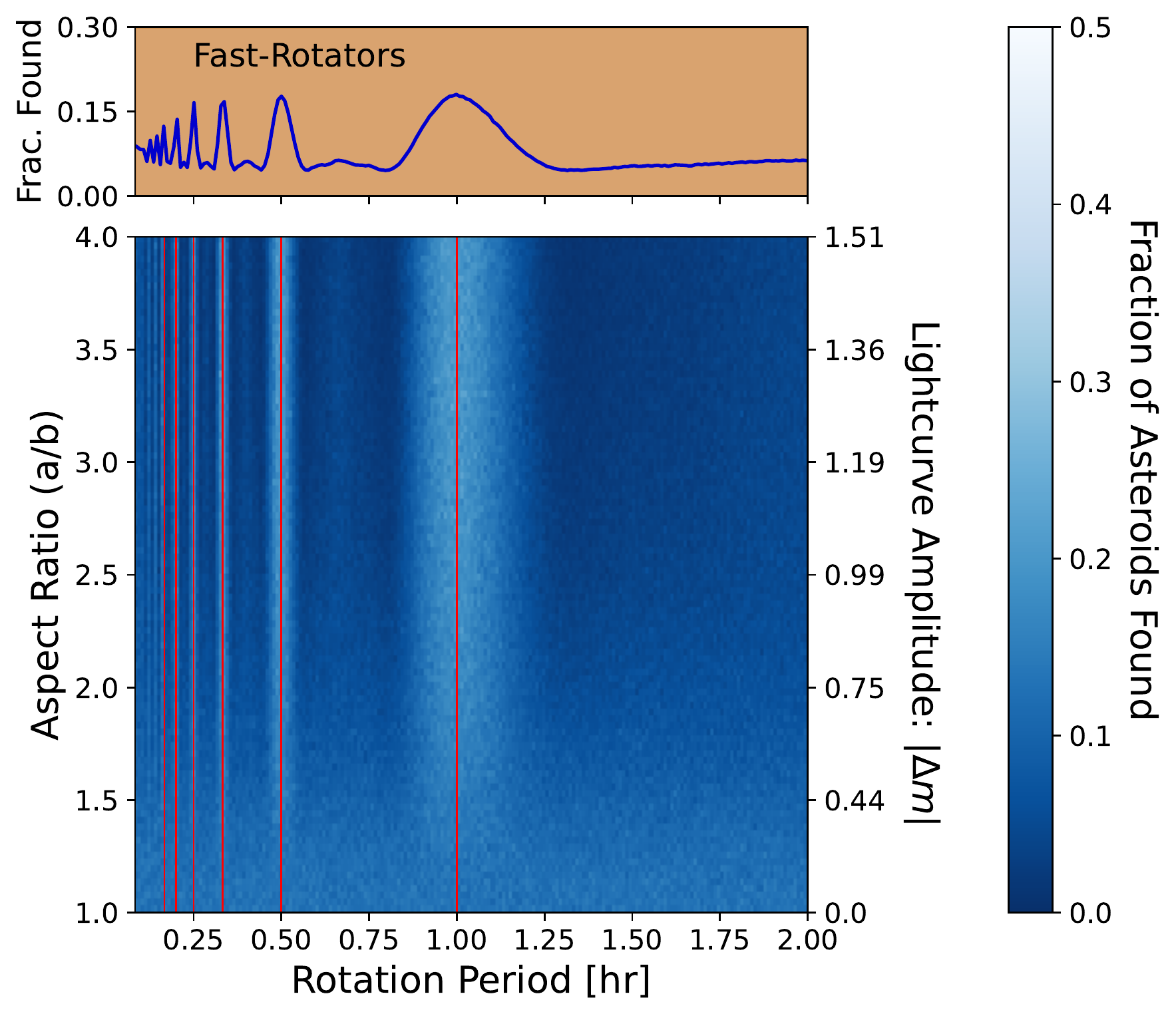}
\caption{Detection statistics of Survey~3 as a function of aspect ratio ($a/b$) and rotation period. Each cell on the 100 x 200 grid corresponds to the injection-recovery test of $10^{3}$ synthetic asteroids, so this heatmap represents $2\times10^{7}$ underlying minor planets. The synthetic asteroids all have $m_r = -0.1$, so these objects were visible to the survey exactly half of the time. From left-to-right, the red vertical lines correspond to periods commensurate with the ten-minute cadence: $P = [10, 12, 15, 20, 30, 60]\,\text{min}$.}
\label{fig:aspectVsPeriod}
\end{figure}

We first consider fast-rotators before expanding our analysis to other $P$. Figure \ref{fig:aspectVsPeriod} provides a heatmap of detection probabilities of $m_r = -0.1$ objects on a grid of $(a/b) \in [1.0, 4.0]$ versus $P \in [5, 120]\,\text{min}$ cells for Survey~3. We keep the same modeling procedure as we did for Figure \ref{fig:aspectVsMag}, but we increase the number of x-axis bins to 200 to increase our $P$ resolution. From the structure of Figure \ref{fig:LCDB} and the representative values in Table \ref{tab:cadenceTable}, we can infer that fast-rotators should have $d < 200\,\text{m}$. For this range, a uniform distribution of $P$ is appropriate. Although Maxwellian rotation rates have been theoretically-predicted and observationally-confirmed for collisionally-evolved asteroids \citep{binzel1988collisionalEvolution}, objects with $d < 40\,\text{km}$ do not conform to this result \citep{pravec2000asteroidrotation}. For example, \cite{pravec2008spinDistribution} found that $3-15\,\text{km}$ diameter Mars-crossing MBAs tend towards a uniform distribution of $P$ due to the YORP effect. Fast-rotators will have smaller diameters than this sample of Mars-crossers, hence our choice of uniform $P$.

Asteroid shape affects the detection probabilities near the system's limiting magnitude when $|\Delta m| \gtrsim w$. For objects with $|m_{r}| \sim w$, randomness from tracklet-building in the low SNR regime drives the probability of detection. Discovery fractions for these low-amplitude lightcurves are not driven by the asteroid's physical parameters but instead by the survey's intrinsic width $w$ at $m_{lim}$. For more elongated objects, we find that detection statistics depend heavily on the rotation period. The most commonly-discovered objects have periods that are commensurate with the survey cadence, corresponding to the spikes on the top panel of Figure \ref{fig:aspectVsPeriod}. If $P$ is nearly an integer ratio with the cadence, then consecutive exposures view the asteroid at nearly the same rotational phase. This selection effect has been identified in past studies \citep{masiero2009thousand} but not explicitly connected to the discovery of high-amplitude lightcurves. Commensurate rotators, in addition to being detected more often, are also more likely to have underestimated $|\Delta m|$ which could skew inferred shape distributions towards smaller axial ratios.

Next, we explore the full range of rotation rates --- $P \in [1\,\text{s}, 1024\,\text{h}]$ --- and all representative survey cadences from Table \ref{tab:cadenceTable}. The $P$ dependence of detectability in Figure \ref{fig:aspectVsPeriod} is strongest for $(a/b) > 2$, so we calculate the discovery fractions versus $P$ for $m_r = -0.1$ asteroids averaged over $(a/b) \in [2, 4]$. Each run sampled $10^{4}$ asteroids in each of 1000 bins evenly-spaced in $\log_{10}(P)$ for a total of $10^{7}$ asteroids per survey. Values of $P$ that are close to integer ratios with the survey cadence have enhanced discovery rates. The fractional range of values, $\Delta P/P$, over which the enhancement occurs decreases as the rotation rate increases. For example, a ten-minute survey cadence will observe a $P \sim 9-11\,\text{min}$ asteroid at $\phi$ within $\pm 10\%$ during consecutive exposures. In contrast, an asteroid with $P \sim 1\,\text{h}$ must have $P \sim 58-62\,\text{min}$ to show the same degree of aliasing.

\begin{figure*}
\epsscale{1.1}
\plotone{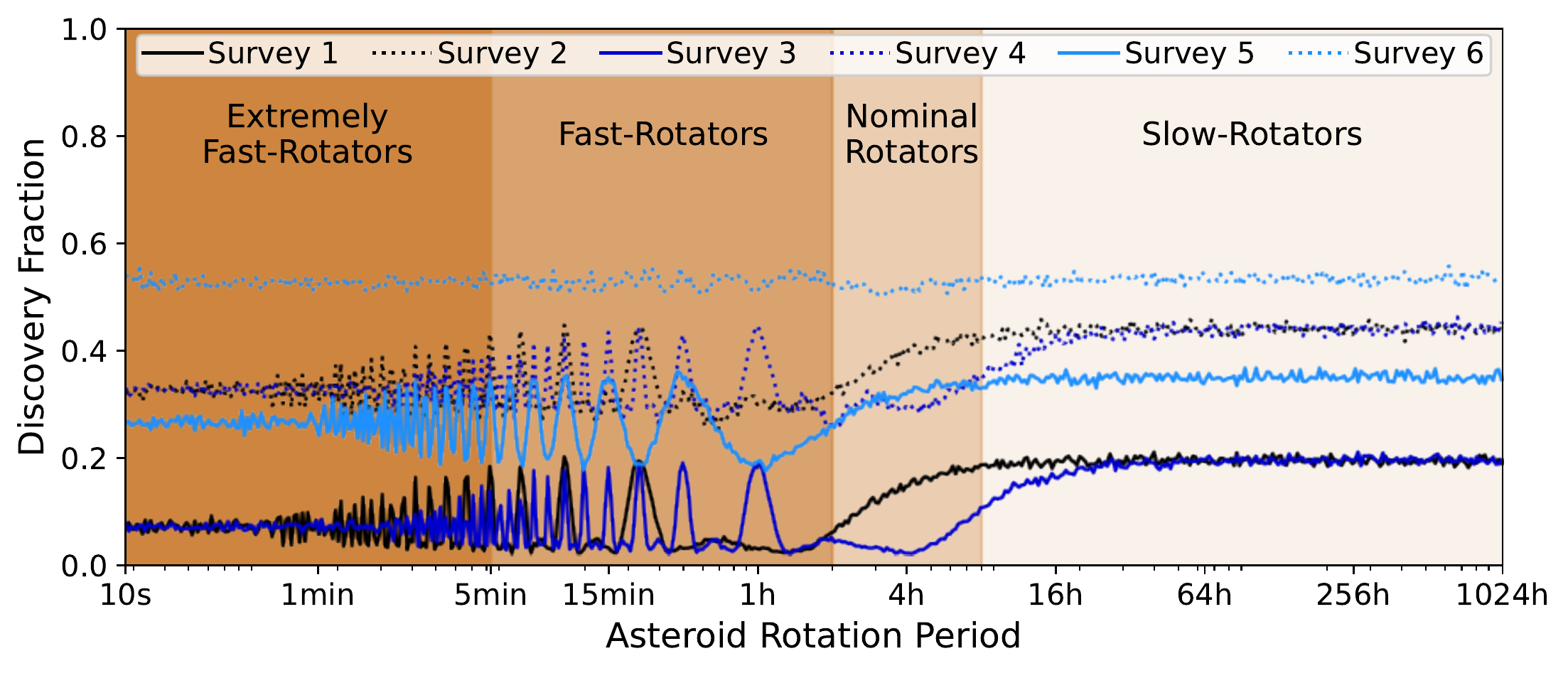}
\caption{Fraction of asteroids detected by each mock survey as a function of rotation period, where the results have been averaged over a uniform distribution of aspect ratios: $(a/b) \in [2, 4]$. Points on the x-axis are equally-spaced in $\log_{10}(P)$.}
\label{fig:aspectVsPeriodAll}
\end{figure*}

In Figure \ref{fig:aspectVsPeriodAll}, the discovery fraction asymptotically converges to limiting cases for extreme $P$. The tracklet-building probability for slow-rotators approaches the fraction of the lightcurve brighter than the system limit (after accounting for the system's $w$). In these cases, the asteroids' rotational phases in subsequent exposures are correlated; whichever $\phi$ is viewed during the first exposure is similar to the $\phi$ for all observations. As we expect, Surveys 1 \& 3 and Surveys 2 \& 4 converge to the same discovery fractions since these pairs have the same tracklet-building criteria in Table \ref{tab:cadenceTable}. For the cadences that we consider, nominal rotators mark the transition from discovery fractions that are dominated by commensurability concerns to the slow-rotator limit.

Extremely fast-rotators are observed at nearly randomized rotation phases in each frame so long as their periods are still longer than the exposure time. Their discovery probability is essentially a series of independent events, which also results in a convergence on Figure \ref{fig:aspectVsPeriodAll} for Surveys 1 \& 3 and Surveys 2 \& 4. This limiting case is also relevant if asteroids are visible for many nights, such as the case considered by \cite{navarroMeza2021lightcurveBias}. These asteroids violate our assumption in Section \ref{sec:model} that $\phi$ does not change while the image is being obtained (i.e. 15s for LSST). The photon flux from extremely fast-rotators is the integrated brightness across the exposure, which could alter their discovery fractions by dampening the measured $|\Delta m|$. We anticipate that this effect will increase the detectability of extremely fast-rotators, but a robust confirmation of our prediction would require testing via survey-specific asteroid extraction pipelines. Since the sample of known extremely fast-rotators is small, further work like that of \cite{beniyama2022video} is needed to characterize the prevalence of these extreme objects.

Overall, NEO surveys are biased towards discovering elongated fast-rotators with periods that are commensurate with the survey cadence. While the LSST-like cadence in Survey~5 performs well, the additional visit for Survey~6 essentially eliminates the dependence on rotation period. In the latter case, the discovery fraction is approximately $0.5$ across the range of examined $P$ values. For real minor planet populations, the exact location of the spin barrier should negligibly affect shape-driven variability as long as the survey cadence is not in the 2-3$\,\text{h}$ range that could be commensurate with $P_{c}$. Nonetheless, future efforts to debias $P$ distributions from survey discoveries must account for the interplay between the spin barrier's location and allowable axial ratios on either side of $P_{c}$.

\begin{figure*}
\epsscale{1.1}
\plotone{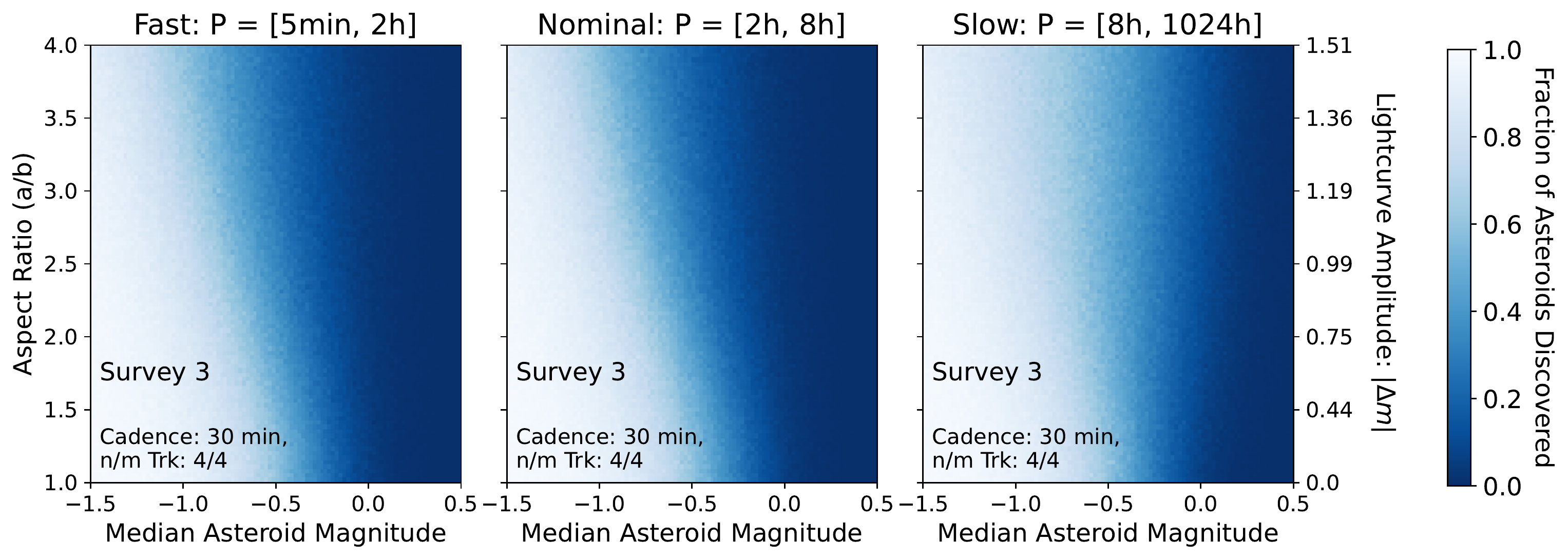}
\caption{Results in $(a/b)$ vs. $m_r$ space from applying Survey~3 to fast-rotators, nominal rotators, and slow-rotator regimes. This format is analogous to the layout in Figure \ref{fig:aspectVsMag} but compares the shape-driven selection effects for different $P$ regimes. Each heatmap represents a 100 x 100 grid, where each cell corresponds to the search for 1000 NEOs.}
\label{fig:Pregimes}
\end{figure*}

Finally, we confirm that recreating Figure \ref{fig:aspectVsMag} with different observed rotation regimes does not lead to a change in our conclusions. Continuing to use Survey~3 as an example, we apply the same computational methods for initializing grids of model asteroids and simulating their detectability. Here, we do three simulations with three different $P$ regimes: uniform within the fast-rotators, nominal rotators, and slow-rotators. Our results (Figure \ref{fig:Pregimes}) show that slow rotators are less sensitive to shape-driven variability, while fast-rotators and nominal rotators exhibit a similar overall degree of selection effects. Nonetheless, elongated objects are selected against in each rotation regime.

\section{Population-Level Ramifications} \label{sec:population}

Thus far, we have established trends in single-object, small body detection efficiencies. Turning to shape-driven variability in population-level statistics, we consider the possibility that debiased shape distributions and SFDs could be incorrect due to the inefficient discovery of elongated objects.

\subsection{Inferring Shape Distributions}

The axial ratios of minor planet populations can illuminate their collisional histories, such as the timing and physics of family-forming impacts. For example, \cite{leinhardt2000collisionSim} used numerical collision models to show that direct erosion from minor impacts decreases $(a/b)$ over time. Others have hypothesized that low-level seismic activity associated with these events indirectly drives older asteroids towards spherical shapes \citep{richardson2004impactSeismic}. Although determining the geometry of an individual small body requires high cadence observations to resolve the lightcurve, the distribution of axial ratios can be determined from sparse, aggregate photometry from a survey \citep{szabo2008shape, masiero2009thousand, cibulkova2018shapes, mcneill2019shapes}.

Past efforts to determine shape distributions have implicitly assumed that the discovery pipeline itself was not sensitive to the axial ratios. Instead, the debiasing procedure focused on the probabilities of inferring $P$ and $|\Delta m|$ of already-discovered asteroids. For example, the amplitudes of lightcurves with $|\Delta m|$ near the survey's photometric noise limit are often difficult to recover \citep{masiero2009thousand}. In the case of kilometer-scale MBAs observed at moderate cadences, such as the study by \cite{masiero2009thousand}, ignoring shape-driven variability could be acceptable. For NEOs probed with sparse (i.e. single-night) photometry, however, shape-driven selection effects may bias estimates of the $(a/b)$ distribution. To demonstrate this point, we initialized a population of $10^{5}$ asteroids with aspect ratios following an exponential distribution

\begin{equation} \label{eq:probAspect}
    f(x) = \bigg(\frac{1}{\beta}\bigg)\exp{\bigg(\frac{-x}{\beta}\bigg)}\,,
\end{equation}

\noindent where $x \equiv (a/b) - 1$ and the scale parameter $\beta$ (or inverse rate parameter) is taken to be unity. For the other inputs to Equation \ref{eq:lightcurve}, we randomized $P \in [4, 8]\,\text{hr}$, $\phi \in [0, 2\pi]$, and $m_r \in [-0.25, 0.35]$. Since newly-discovered NEOs are almost always smaller in physical size than the aforementioned YORP-affected Mars-crossers from \cite{pravec2008spinDistribution}, we use a uniform distribution of rotation rates instead of the canonical Maxwellian distribution. Nonetheless, the results of Section \ref{sec:results} on rotation rates show that the $P$ distributions should not affect population-level results.

Our assumed underlying axial ratios are not necessarily representative for any specific set of minor planets. Instead, we use this fiducial model to illustrate shape-driven variability because Equation \ref{eq:probAspect} is conveniently dictated by one free parameter: the steepness of the exponential decay. Through injection-recovery exercises to determine the bias in $\beta$, our methods will demonstrate that selection effects against high-amplitude lightcurves elucidated in Section \ref{sec:results} must be considered when attempting to determine $f(x)$ from observational data. Determining the actual functional form of NEO shapes will require detailed simulations of survey-specific findings that account for the systematic tendency to miss high-amplitude lightcurves.

Previous work by \cite{masiero2009thousand}, the Thousand Asteroid Lightcurve Survey, published MBAs shapes that were best-fit by a piece-wise quadratic. However, that study neither considered an exponential form as a candidate distribution nor accounted for undiscovered high $|\Delta m|$ objects in their debiasing procedure. Moreover, the asteroids under consideration by \cite{masiero2009thousand} were both an order-of-magnitude larger than typical NEOs and observed for multiple nights. Finally, the NEOs that we consider may also exhibit morphologies modified by close planetary encounters. Taken together, these caveats mean that the shape distribution derived by that research is likely inapplicable to our NEO-specific situation.

Recent commentary on the NEO shape distribution by \cite{thirouin2018MANOS} reported that the number of elongated ($(a/b) \gtrsim 2$) asteroids is comparable to the number of more spherical objects. Because shape-driven discovery effects are ignored in that research, our Section \ref{sec:results} suggests that this claim provides a lower bound on the degree of asphericity in NEOs. The cumulative distribution of Equation \ref{eq:probAspect} shows that $\sim63\%$ of our assumed shapes have $(a/b) < 2$, comparable to the fraction given by \cite{thirouin2018MANOS}.

\begin{figure}
\epsscale{1.1}
\plotone{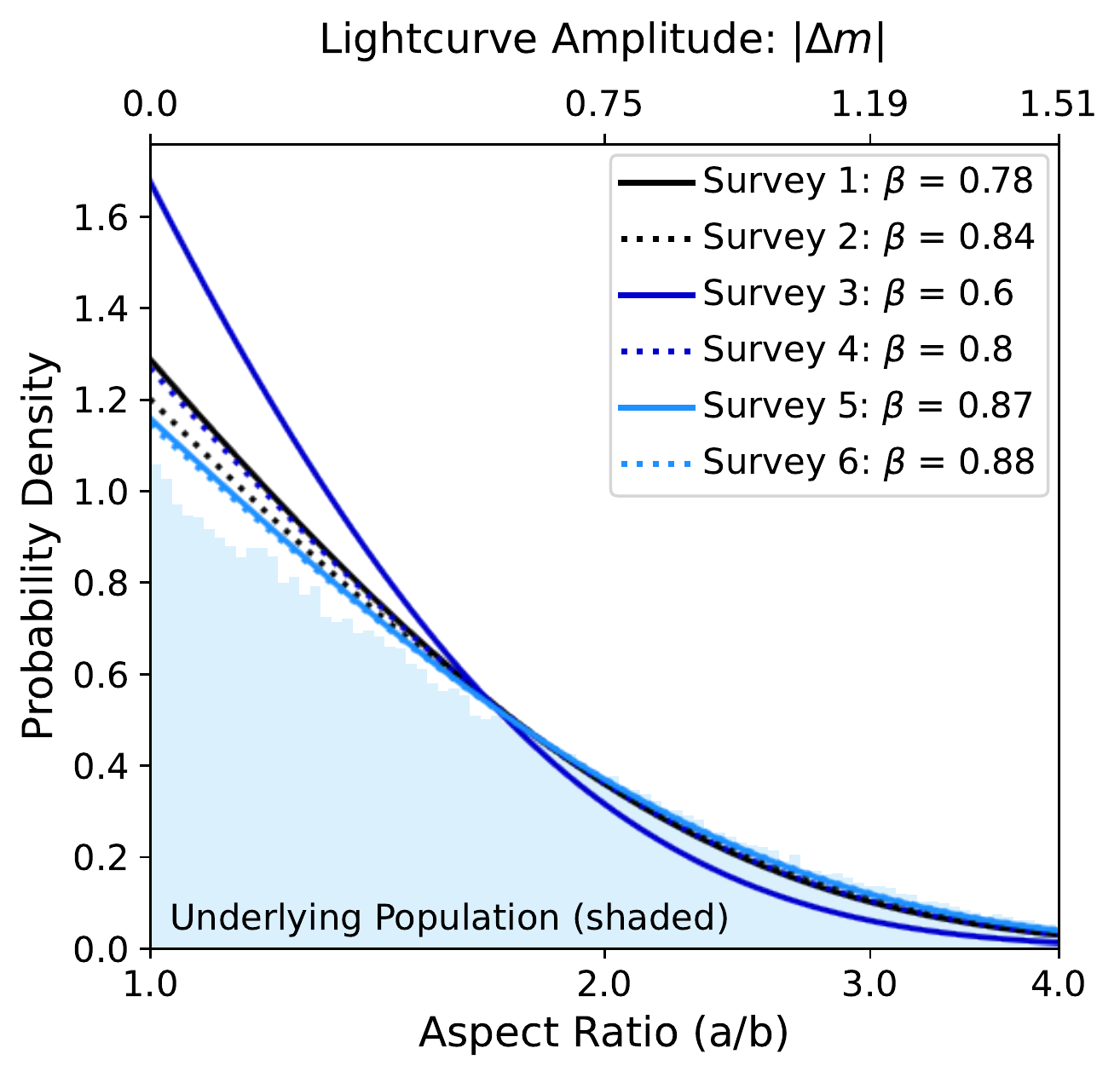}
\caption{Results of investigating shape-driven selection effects on simulated small bodies with underlying axial ratios given by the shaded background. The recovered scale parameter (from Equation \ref{eq:probAspect}) of the population of identified synthetic objects from each survey is shown as an annotation in the legend for each survey.}
\label{fig:shapeDist}
\end{figure}

With these synthetic small body parameters, we ran each of the mock surveys in Table \ref{tab:cadenceTable} with the aforementioned methods for generating lightcurves to model the discovery of the synthetic asteroids. Regardless of $|\Delta m|$, we assumed that $(a/b)$ was determined if that object was discovered by the survey. We then fit an exponential distribution to the asteroids that were identified by our simulation to compare with the underlying population from Equation \ref{eq:probAspect}. Figure \ref{fig:shapeDist} shows normalized probability density functions of the underlying axial ratios and recovered $(a/b)$ distributions for each injection-recovery survey run.

As expected from our results in Section \ref{sec:results}, we find that our simulations miss many objects that are severely elongated. In turn, the scale parameter $\beta$ in Equation \ref{eq:probAspect} is overestimated for all strategies and inconsistent between the strategies themselves. Survey~3, with the longest time between observations and most severe detection threshold, is the most hampered by elongation and correspondingly returns the most incorrect observed shape distribution. Of the odd-numbered surveys, the ones that require asteroids to be visible in all exposures to be identified, the LSST-like survey performs the best. This result is a consequence of LSST's short overall temporal baseline with only two exposures compared to the four exposures by the Pan-STARRS analog. We stress that our results are sensitive to both cadence and discovery criteria, and are only relevant to objects with average brightness near the survey field's limiting magnitudes. Nonetheless, the most common apparent magnitude for newly-discovered minor planets is $m_r \sim 0$.

\subsection{Inferring Size-Frequency Distributions}

To this point, our analysis has remained agnostic to asteroid size by considering apparent magnitude, $m_r$, relative to the survey's limit. From the previous exercises, we estimated discovery probabilities for individual objects of a given variability and for underlying shape distributions of objects near a given survey's limiting magnitude. Another insightful characterization of small bodies is their SFD, which can illuminate their collisional histories \citep{harris2015populationNEA, granvik2018NEOdebiased} and provide robust estimations of planetary impact rates \citep{bottke2000NEOscience, heinze2021neoPopulation}. Here, we demonstrate that deducing correct SFDs may require accounting for shape-driven variability if the axial ratios and sizes of NEOs are correlated. Specifically, we conduct injection-recovery tests on fiducial populations of asteroids with underlying SFDs that are explicitly connected to their $(a/b)$ distribution.

Objects with $d \gtrsim 200\,\text{km}$ are expected to have nearly spherical shapes due to hydrostatic balance, a trend which is confirmed by noting the small $|\Delta m|$ of asteroids in Figure \ref{fig:LCDB} beyond the $10^{2}\,\text{km}$ scale. This regime only applies to some MBAs and TNOs; the shapes of NEOs are dictated by other considerations. The correlation, if any, between axial ratio and size for smaller objects is currently unknown. Some observations have interrogated this possible relationship, but only within a set of already-known asteroids. For example, \cite{hatch2015smallNEAs} and \citep{thirouin2016MANOSphotometry} each reported a weak tendency for elongation to positively correlate with increasing size. However, the lack of high ($a/b$) at small $d$ may actually indicate a discovery bias against elongated asteroids. Without rigorous debiasing specific to shape-driven lightcurve variability, the true correlation (if any) between ($a/b$) and $d$ will remain elusive.

The \textit{a priori} expectation for a shape-size relationship is also unclear, as competing processes push the trend oppositely. One could hypothesize that NEOs with smaller diameters are more likely to be elongated due to their lower surface gravity \citep{veres2017lsstSimulation}. Alternatively, a size-dependent tensile strength \citep{holsapple2007spinLimits} could manifest itself through size-shape correlations. Objects with $d \sim 1\,\text{km}$ and smaller are affected by YORP forces, which can spin-up asteroids to increase their aspect ratios or destroy them entirely \citep{vokrouhlicky2002YORP}.

Given this uncertain size-shape trend, we arbitrarily take the axial ratios of small bodies to be negatively correlated with their sizes for this exercise. First, we generate an underlying asteroid population that follows a power-law SFD

\begin{equation} \label{eq:sfd}
    N(>d) \propto d^{\alpha+1}\,,
\end{equation}

where $d$ represents physical size, $\alpha$ is the slope of the power-law, and $N(>d)$ is the number of asteroids greater than size $d$. We assume $\alpha = -3.5$, a characteristic observational result for small body populations that also emerges naturally from first-principles for an evolved, self-similar system \citep{dohnanyi1969collisionalModel}.

From this SFD, we can calculate the number of asteroids between two sizes $d_{1}$ and $d_{2}$ as

\begin{equation} \label{eq:sfdDiff}
    N(d_{1} > d > d_{2}) = N(>d_{2})\bigg(1-\bigg(\frac{d_{1}}{d_{2}}\bigg)^{\alpha+1}\bigg)\,,
\end{equation}

\noindent where $d_{1} > d_{2}$, so $N(>d_{2}) > N(>d_{1})$ for $\alpha < 0$.

\begin{figure*}
\epsscale{1.05}
\plotone{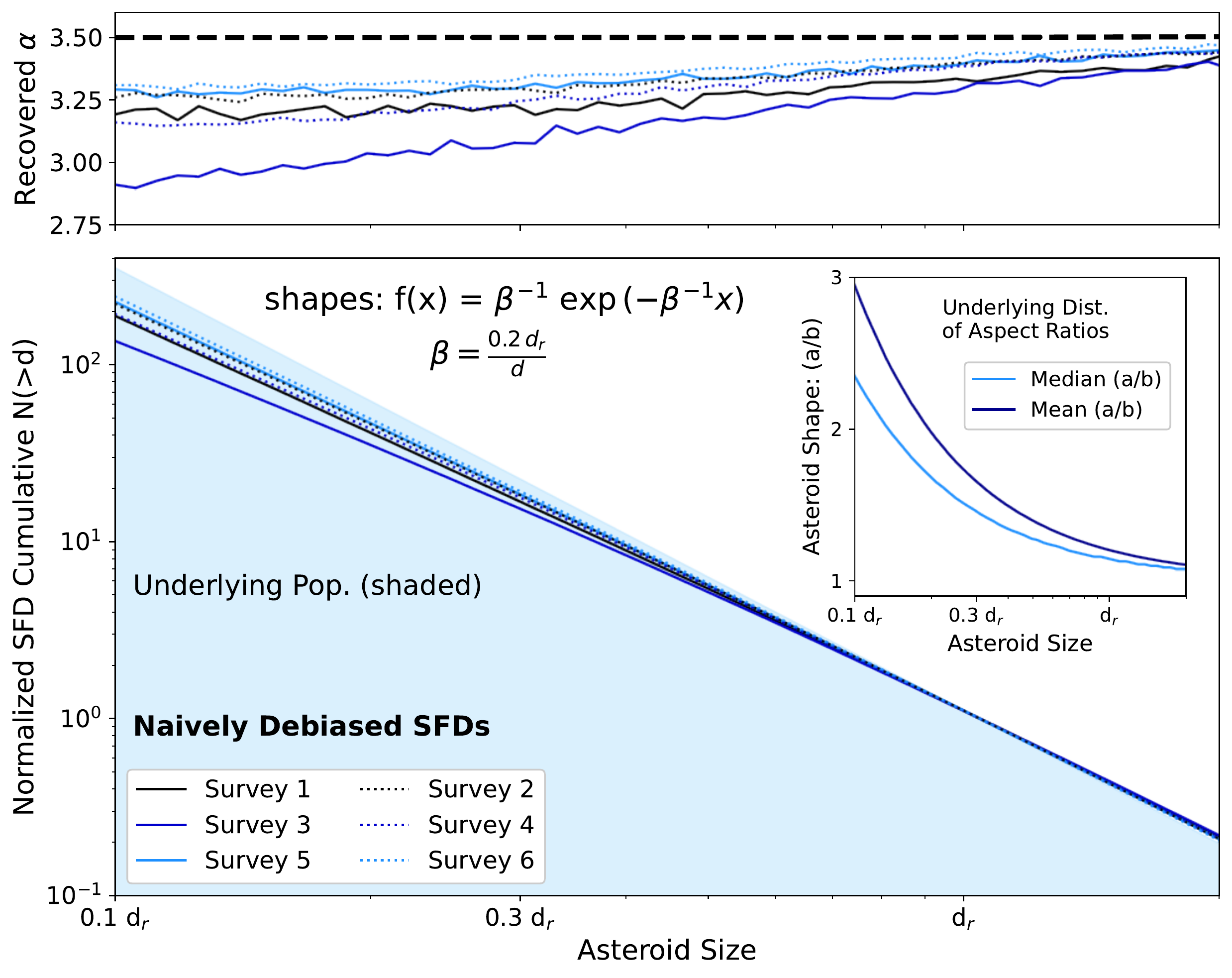}
\caption{Underlying populations of synthetic and recovered asteroids from our model survey, shown as the distribution of counts versus asteroid size. We display the y-axis in terms of the cumulative number of objects larger than the given asteroid size. The x-axis gives the asteroid sizes in terms of $d_{r}$, the size for which the na\"ive debiasing procedure was referenced.}
\label{fig:sfdSurvey}
\end{figure*}

For this exercise, we initialize $n_{\text{bins}} = 100$ bins uniformly spaced in $\log_{10}(d)$ and calculate $N(d_{i+1} > d > d_{i})$. We indexed the $i^{\text{th}}$ SFD size bin with endpoints $a_{i}$ and $a_{i+1}$. Since power-law SFDs are scale-free, we range the asteroid sizes with respect to a reference size $d_{r}$ such that $d \in [0.1\,d_{r}, 30\,d_{r}]$. This normalization continues our study's agnosticism towards specific values for a survey limiting magnitude and small body sizes. We plot the underlying distribution and the $\alpha = 3.5$ value on the bottom and top panels of Figure \ref{fig:sfdSurvey}, respectively.

In each size bin, we generate $10^{5}$ synthetic asteroids with uniformly distributed $P \in [2, 8]\,\text{h}$, $m_r \in [-0.5, 0.5]\,\text{mag}$, $\phi \in [0, 2\pi]$, and physical size $d \in [d_{i}, d_{i+1}]$. These bin widths are sufficiently small that a uniform distribution within each bin is fine. Next, we assign each of these asteroids an aspect ratio in accordance with Equation \ref{eq:probAspect} with $\beta = (0.2\,d_{r})/d$ such that smaller asteroids are typically more elongated. The size-dependent mean and median shape are plotted on the inset of Figure \ref{fig:sfdSurvey}. We conservatively chose $\beta$ such that the median axial ratio approached $(a/b) \sim 2$ for small objects in our SFD (Figure \ref{fig:sfdSurvey}). This fiducial benchmark is based on the reported distribution by \cite{thirouin2018MANOS} of approximately equal numbers of spherical and $(a/b) \sim 2$ NEOs.

We next simulate all surveys from Table \ref{tab:cadenceTable} on the asteroid populations that were initialized in each size bin. From this procedure, we obtain the fraction of objects found by the $j^{\text{th}}$ survey in a given size bin: $f_{j}(d_{i+1} > d > d_{i})$. Next, we calculate the number of asteroids in the SFD that would have been found in each size bin as $N_{\text{f}, j}(d_{i+1} > d > d_{i}) = f_{j}(d_{i+1} > d > d_{i}) \times N(d_{i+1} > d > d_{i})$. At this point, we have a set of ``observed objects" for each survey in each size bin for which we take the cumulative sums of objects in bins larger than $d_{i}$ to determine $N_{\text{f}, j}(>d_{i})$. We then perform a na\"ive debiasing procedure that ignores the correlation between $(a/b)$ and diameter. We normalize all recovered SFDs such that the debiased cumulative counts $N_{\text{d}, j}(>d_{r}) = 1$ at the reference size $d_{r}$ and do the same for the underlying set of asteroids. By dividing the values in each $N_{\text{f}, j}(>d_{i})$ bin by the same survey-specific normalization factor, we derive the ``debiased" SFD $N_{\text{d}, j}(>d)$ assuming that the underlying distribution of axial ratios is constant across all sizes with $d_{r}$ as a reference.

We plot the resulting incorrectly debiased SFDs, $N_{\text{d}, j}(>d)$, on the bottom panel of Figure \ref{fig:sfdSurvey}. Additionally, we solve for the inferred $\alpha$ values in sequential bins from the normalized $N_{d, j}$ and corresponding $d_{i}$ (plotted in top panel of Figure \ref{fig:sfdSurvey}. Because the discovery fraction is overestimated at small sizes and underestimated at large sizes, the recovered $\alpha$ is systematically lower for all simulated surveys across the entire size range. The shape distribution is more extreme at smaller sizes, so estimated $\alpha$ values diverge farther from the true value for $d < d_{r}$. These effects are most pronounced for Survey~3, which has the strongest shape-driven discovery bias. Had we reversed the direction of the underlying size-shape correlation, our simulations would have overestimated $\alpha$ instead.

Although the parameters applied to Equations \ref{eq:probAspect} \& \ref{eq:sfd} do not correspond to a specific set of NEOs, we intend for this exercise to demonstrate the possible ramifications of omitting shape-driven variability when debiasing systematic survey yields. Our procedure does not account for errors in observed $H$ from observing lightcurves at their maximum, but Figure \ref{fig:Hmag} show that this consideration can be safely ignored. To determine the actual parameter space affected by shape-driven variability, the underlying asteroid size-shape trend must be constrained through forward-modeling simulations that also account for shape-driven discovery biases. We anticipate that published SFDs are more likely to be reliable at sizes for which observed NEO populations are more complete. For example, completeness quickly rises above $30\%$ for $d \gtrsim 200\,\text{m}$ \citep{tricarico2017NEOcompleteness}. We have taken all asteroids in the population to have been observed only near $m_{lim}$, but this assumption is increasingly violated for increasing physical size.

\section{Discussion \& Implications} \label{sec:discussion}

We have implemented a basic model of small body lightcurve variability, examined fiducial survey strategies, and assessed the impact of shape-driven selection effects. Now, we remark on the relevance of our work to discrepancies in published SFDs for NEOs, discuss the broader applicability of our work to small body populations besides NEOs, and provide actionable recommendations for next-generation initiatives like LSST. Throughout this commentary, we mention directions for future work that could further illuminate shape-driven selection effects.

\subsection{Potential Biases in NEO Statistics}

Although the census of NEOs whose impacts would be catastrophic to Earth's biosphere is nearly complete, smaller asteroids ($a \sim 100\,\text{m}$) are less known. \cite{harris2015populationNEA} derived a widely-cited SFD for $d \gtrsim 3\,\text{m}$ based on calibrating forward-model inputs to the fraction of re-detected objects in historical survey data. Results from some recent work \citep{granvik2018NEOdebiased, heinze2021neoPopulation}, however, have diverged from \cite{harris2015populationNEA} by discovering fewer objects than predicted at small diameter. This discrepancy is often attributed to trailing loss, where high $H$ objects must be close to Earth to be sufficiently bright as to be identified. Correspondingly, these asteroids exhibit rapid apparent motion and may have high enough angular velocities to trail across multiple pixels during the exposure. As such, the photons collected during the exposure are spread over multiple pixels and may not be efficiently detected and/or properly accounted in the debiasing procedure.

While this trailing selection effect is a viable explanation, our results suggest that lightcurve variability could also affect SFDs should a correlation between size and shape exist in the discrepant region of parameter space. Our results in Section \ref{sec:population} indicate that the shape distribution of NEOs may not be properly debiased. Although high-cadence photometric follow-up can determine the axial ratios of individual objects, these asteroids must first be identified as candidate minor planets by wide-field surveys that can impose shape-dependent biases. Thus, a careful injection-recovery test of elongated NEOs for existing and future surveys must be performed to assess the degree to which asteroids experience adverse selection effects related to their $|\Delta m|$.

While we focus on shape-driven effects that dominate most asteroidal variability (including \om's lightcurve; \cite{Meech2017}), asymmetric surface albedos can also cause brightness variability \citep{mashchenko2019modelling}. Thus, albedos that vary across the minor planet's surface could also introduce discovery biases. This issue is less amenable to simulation without an analytic lightcurve analog to Equation \ref{eq:lightcurve}, so we leave the question of detection efficiencies for objects with albedo-driven variability to future work. Nonetheless, our simulations provide insight towards small bodies of varying albedo since the rotation period sets the variability timescale; highly-variable objects should be less detectable, even if those variations are attributed to mineralogy instead of axial ratio.

In either the case of shape-driven or albedo-driven variability, lightcurve amplitudes may be underestimated if the minima are unobservable \citep{navarroMeza2021lightcurveBias}. Estimated absolute magnitudes may also be affected by the same variability bias. Together, these effects could hamper efforts to determine SFDs. Therefore, future debiased NEO populations from observational data should include a comprehensive treatment of lightcurve variability.

\subsection{Applicability to Interstellar Interlopers}

In the current era of survey sophistication, our simulations are necessarily more relevant to NEOs than ISOs. Around the epoch of its discovery\footnote{\url{https://minorplanetcenter.net//mpec/K17/K17UI1.html}}, \om's lightcurve minima were approximately 21$^{\text{st}}$ magnitude and close to the Pan-STARRS system limit. Although \om{} was visible at all rotational phases to Pan-STARRS at this time, it's possible that this first ISO could have been detected earlier if its axial ratio were smaller. Without a statistically-significant population of \om-like objects detected near the survey limit, it is infeasible to apply this study's results to revise estimates of the Galactic ISO reservoir. Fortunately, the LSST may discover dozens of interlopers that are similar in size to \om{} \citep{cook2016realistic, levine2021exotic, hoover2022LSST}. If these objects also display dramatic lightcurve variability, then considerations on shape-driven bias will have to be incorporated into population-level estimates.

Because ISOs are likely to have short observational windows, rapid follow-up is required to capitalize on any discovery. Even if future elongated interlopers do become substantially brighter than the limiting magnitudes of the discovery observatories, early identification when the objects' brightness straddles the limit of detectability would pay scientific dividends. Specifically, surveys that reduce systematic bias against elongated ISOs would permit more precise orbital fitting, longer baselines to search for cometary activity, and a higher likelihood to execute an intercept mission \citep{Seligman2018}.

\subsection{Optimized Survey Strategies}

In all six of our fiducial strategies in Table \ref{tab:cadenceTable}, high-amplitude lightcurves that were visible at most times went unidentified. Since generating tracklets requires multiple detections of the same object on the same night, accounting for shape-driven effects must be done for any survey to improve the derived SFD and rotation period distribution. As we found that more lenient detection criteria can lead to a marked increased in finding elongated asteroids, one option to identify more high-amplitude lightcurves would be to construct tracklets from fewer points than the nominal detection criterion. Then, forced photometry could be performed at the hypothesized locations of asteriods in images where the objects are not found by the basic pipeline. These searches could be computationally expensive and encounter numerous false positives, but precovering specific elongated objects in this manner could be feasible and scientifically justifiable. Nonetheless, extending asteroid identification pipeline capabilities should be considered holistically in the context of the survey's goals.

Our Surveys 1-4 can provide insight on past and ongoing NEO searches, and we chose the parameters of Surveys 5 \& 6 to compare these already-designed surveys with a likely LSST cadence. In our simplified model, we found that the LSST's cadence compared favorably on a nightly basis to the other strategies in Table \ref{tab:cadenceTable}. Nevertheless, many small bodies will be difficult to followup due to the LSST's faint limiting magnitude. Therefore, it's imperative to understand the aggregate detection statistics from LSST itself. In anticipation of this expanded catalog of minor planets, we hope that this study motivates future efforts to implement rigorous shape debiasing procedures towards a more complete understanding of asteroid populations and collisional evolution.

One example of a non-uniform cadence that might affect some shape-driven discovery biases is a ``last opportunity" revisit for some fields that were observed towards the beginning of the night. Some period-driven commensurability may be alleviated if the final visit time is not an integer multiple of the initial cadence. To examine this possibility, we compare the uniform cadence of Survey~3 with a non-uniform survey with observations at $t = \{0, 30, 60, 337\}\,\text{min}$ where the final time was chosen as a prime number that is spaced far from the other three observations. By keeping n/m tracklet to 4/4, we construct a simulation similar to Survey~3 with a long baseline, non-commensurate revisit.

We keep the asteroid parameter distributions that were used to generate Figure \ref{fig:aspectVsMag} and compare the results by subtracting discovery probabilities in each grid cell. In all cases, the absolute difference in discovery fraction between the two surveys is lower than 10\%. Neither survey outperforms the other one, and we do not include the resulting heatmap in this manuscript since there are no obvious trends to visualize. Because our imposed non-uniform cadence requires that the NEO remains above the survey's altitude limit for $>5.5\,\text{h}$, this strategy constrains the field of regard. Nonetheless, these results are indicative of the minor changes that could be expected from non-uniform and long-baseline cadences.

\section{Summary \& Conclusion} \label{sec:conclusion}

With this study, we have considered the consequences of shape-driven lightcurve variability on the detection efficiencies of small bodies in systematic observational campaigns. Our simulations demonstrate that missing elongated objects in systematic asteroid surveys is unavoidable near the system's limiting magnitude. Due to shape-driven lightcurve variability, identifying a given aspherical body is often a matter of luck. Objects with extreme lightcurves and median brightness near a survey's detection limit suffer the most, but asteroids with modest axial ratios can be missed by surveys. Inferred shape distributions and SFDs, important tracers of dynamical and collisional histories, may be affected by high-amplitude lightcurves.

Shape-driven selection effects are dampened due to the inherent width in survey efficiency functions near the limiting magnitude. Because $m_r \sim 0$ asteroids are observed at low $\text{SNR}$, the probability of a given elongated small body being visible in an image is driven by both its rotational phase and a random variable based on the width factor $w$ in Equation \ref{eq:probability}. If $w$ were made smaller by requiring a larger limiting per-detection SNR, then asteroid axial ratios would factor even more into their discoverability. 

The population-level statistics of successful tracklet creation can be understood through general numerical models like we have performed, or through injection-recovery tests in survey-specific simulators. In the imminent LSST era, the sample of known asteroids will grow by an order-of-magnitude \citep{ivezic2019lsst}; this sample will undoubedtly include some extreme shapes. Additionally, high-cadence lightcurves will continue to be obtained from both amateur efforts and professional followup programs to constrain the axial ratios of individual small bodies. We anticipate that these detailed observations can be applied in conjunction with debiased shape distributions of asteroid families to better understand the collisional history, shape evolution, and dynamics of asteroids.

\vspace{5mm}

\acknowledgments
We thank Greg Laughlin, Aster Taylor, Quanzhi Ye, Sam Cabot, Darryl Seligman, Matt Holman, Larry Denneau, and Emma Louden for useful discussions. WGL acknowledges support from the Department of Defense's National Defense Science \& Engineering Graduate (NDSEG) Fellowship and the Connecticut Space Grant Graduate Student Fellowship (award number P-1776). Our simulations and data analysis used the numpy \citep{harris2020numpy}, scipy \citep{virtanen2020scipy}, and matplotlib \citep{hunter2007matplotlib} packages.

\pagebreak
\bibliography{bibliography}
\bibliographystyle{aasjournal}

\end{document}